 \documentclass[prl,twocolumn,showpacs,amssymb,superscriptaddress]{revtex4}

\usepackage{epsfig}  
\usepackage{graphicx}
\usepackage{dcolumn} 
\usepackage{bm}      
\usepackage[dvips]{color}

\begin{document}
\date{June 10, 2005}

\title{Upper Limit on the Branching Ratio for 
    the Decay $\pi^0 \rightarrow \nu \overline{\nu}$}

\author{A.V.~Artamonov}\affiliation{Institute for High Energy Physics, Protvino, Moscow Region, 142 280, Russia}
\author{B.~Bassalleck}\affiliation{Department of Physics and Astronomy, University of New Mexico, Albuquerque, NM 87131, USA}
\author{B.~Bhuyan}\altaffiliation{Also at the Department of Physics and Astrophysics, University of Delhi, Delhi 110007, India} \altaffiliation{Present address: Department of Physics and Astronomy, University of Victoria, Victoria, British Columbia, Canada V8W 3P6.} \affiliation{Brookhaven National Laboratory, Upton, NY 11973, USA}
\author{E.W.~Blackmore}\affiliation{TRIUMF, 4004 Wesbrook Mall, Vancouver, British Columbia, Canada V6T 2A3}
\author{D.A.~Bryman} \affiliation{Department of Physics and Astronomy, University of British Columbia, Vancouver, British Columbia, Canada V6T 1Z1}
\author{S.~Chen}\altaffiliation{Present address: Department of Engineering Physics, Tsinghua University, Beijing 100084, P.R. China} \affiliation{TRIUMF, 4004 Wesbrook Mall, Vancouver, British Columbia, Canada V6T 2A3} 
\author{I-H.~Chiang} \affiliation{Brookhaven National Laboratory, Upton, NY 11973, USA}
\author{I.-A.~Christidi} \affiliation{Department of Physics and Astronomy, Stony Brook University, Stony Brook, NY 11794, USA}
\author{P.S.~Cooper}\affiliation{Fermi National Accelerator Laboratory, Batavia, IL 60510, USA}
\author{M.V.~Diwan} \affiliation{Brookhaven National Laboratory, Upton, NY 11973, USA}
\author{J.S.~Frank} \affiliation{Brookhaven National Laboratory, Upton, NY 11973, USA}
\author{T.~Fujiwara}\affiliation{Department of Physics, Kyoto University, Sakyo-ku, Kyoto 606-8502, Japan}
\author{J.~Hu} \affiliation{TRIUMF, 4004 Wesbrook Mall, Vancouver, British Columbia, Canada V6T 2A3}
\author{D.E.~Jaffe} \affiliation{Brookhaven National Laboratory, Upton, NY 11973, USA}
\author{S.~Kabe} \affiliation{High Energy Accelerator Research Organization (KEK), Oho, Tsukuba, Ibaraki 305-0801, Japan}
\author{S.H.~Kettell} \affiliation{Brookhaven National Laboratory, Upton, NY 11973, USA}
\author{M.M.~Khabibullin}\affiliation{Institute for Nuclear Research RAS, 60 October Revolution Pr. 7a, 117312 Moscow, Russia}
\author{A.N.~Khotjantsev}\affiliation{Institute for Nuclear Research RAS, 60 October Revolution Pr. 7a, 117312 Moscow, Russia}
\author{P.~Kitching}\altaffiliation{Present address: TRIUMF, Canada.} \affiliation{Centre for Subatomic Research, University of Alberta, Edmonton, Canada T6G 2N5}
\author{M.~Kobayashi} \affiliation{High Energy Accelerator Research Organization (KEK), Oho, Tsukuba, Ibaraki 305-0801, Japan}
\author{T.K.~Komatsubara} \affiliation{High Energy Accelerator Research Organization (KEK), Oho, Tsukuba, Ibaraki 305-0801, Japan}
\author{A.~Konaka} \affiliation{TRIUMF, 4004 Wesbrook Mall, Vancouver, British Columbia, Canada V6T 2A3}
\author{A.P.~Kozhevnikov}\affiliation{Institute for High Energy Physics, Protvino, Moscow Region, 142 280, Russia}
\author{Yu.G.~Kudenko}\affiliation{Institute for Nuclear Research RAS, 60 October Revolution Pr. 7a, 117312 Moscow, Russia} 
\author{A.~Kushnirenko} \altaffiliation{Present address: Institute for High Energy Physics, Protvino, Russia.}  \affiliation{Fermi National Accelerator Laboratory, Batavia, IL 60510, USA} 
\author{L.G.~Landsberg}\affiliation{Institute for High Energy Physics, Protvino, Moscow Region, 142 280, Russia}
\author{B.~Lewis}\affiliation{Department of Physics and Astronomy, University of New Mexico, Albuquerque, NM 87131, USA}
\author{K.K.~Li}\affiliation{Brookhaven National Laboratory, Upton, NY 11973, USA}
\author{L.S.~Littenberg} \affiliation{Brookhaven National Laboratory, Upton, NY 11973, USA}
\author{J.A.~Macdonald} \altaffiliation{Deceased.} \affiliation{TRIUMF, 4004 Wesbrook Mall, Vancouver, British Columbia, Canada V6T 2A3}
\author{J.~Mildenberger} \affiliation{TRIUMF, 4004 Wesbrook Mall, Vancouver, British Columbia, Canada V6T 2A3}
\author{O.V.~Mineev}\affiliation{Institute for Nuclear Research RAS, 60 October Revolution Pr. 7a, 117312 Moscow, Russia}
\author{M. Miyajima} \affiliation{Department of Applied Physics, Fukui University, 3-9-1 Bunkyo, Fukui, Fukui 910-8507, Japan}
\author{K.~Mizouchi}\affiliation{Department of Physics, Kyoto University, Sakyo-ku, Kyoto 606-8502, Japan}
\author{V.A.~Mukhin}\affiliation{Institute for High Energy Physics, Protvino, Moscow Region, 142 280, Russia}
\author{N.~Muramatsu} \affiliation{Research Center for Nuclear Physics, Osaka University, 10-1 Mihogaoka, Ibaraki, Osaka 567-0047, Japan}
\author{T.~Nakano}\affiliation{Research Center for Nuclear Physics, Osaka University, 10-1 Mihogaoka, Ibaraki, Osaka 567-0047, Japan}
\author{M.~Nomachi}\affiliation{Laboratory of Nuclear Studies, Osaka University, 1-1 Machikaneyama, Toyonaka, Osaka 560-0043, Japan}
\author{T.~Nomura}\affiliation{Department of Physics, Kyoto University, Sakyo-ku, Kyoto 606-8502, Japan}
\author{T.~Numao} \affiliation{TRIUMF, 4004 Wesbrook Mall, Vancouver, British Columbia, Canada V6T 2A3}
\author{V.F.~Obraztsov}\affiliation{Institute for High Energy Physics, Protvino, Moscow Region, 142 280, Russia}

\author{K.~Omata}\affiliation{High Energy Accelerator Research Organization (KEK), Oho, Tsukuba, Ibaraki 305-0801, Japan}
\author{D.I.~Patalakha}\affiliation{Institute for High Energy Physics, Protvino, Moscow Region, 142 280, Russia}
\author{S.V.~Petrenko}\affiliation{Institute for High Energy Physics, Protvino, Moscow Region, 142 280, Russia}
\author{R.~Poutissou} \affiliation{TRIUMF, 4004 Wesbrook Mall, Vancouver, British Columbia, Canada V6T 2A3}
\author{E.J.~Ramberg}\affiliation{Fermi National Accelerator Laboratory, Batavia, IL 60510, USA}
\author{G.~Redlinger} \affiliation{Brookhaven National Laboratory, Upton, NY 11973, USA}
\author{T.~Sato} \affiliation{High Energy Accelerator Research Organization (KEK), Oho, Tsukuba, Ibaraki 305-0801, Japan}
\author{T.~Sekiguchi}\affiliation{High Energy Accelerator Research Organization (KEK), Oho, Tsukuba, Ibaraki 305-0801, Japan}
\author{T.~Shinkawa} \affiliation{Department of Applied Physics, National Defense Academy, Yokosuka, Kanagawa 239-8686, Japan}
\author{R.C.~Strand} \affiliation{Brookhaven National Laboratory, Upton, NY 11973, USA}
\author{S.~Sugimoto} \affiliation{High Energy Accelerator Research Organization (KEK), Oho, Tsukuba, Ibaraki 305-0801, Japan}
\author{Y.~Tamagawa} \affiliation{Department of Applied Physics, Fukui University, 3-9-1 Bunkyo, Fukui, Fukui 910-8507, Japan}
\author{R.~Tschirhart}\affiliation{Fermi National Accelerator Laboratory, Batavia, IL 60510, USA}
\author{T.~Tsunemi} \altaffiliation{Present address: Research Center for Nuclear Physics, Osaka University, Japan.} \affiliation{High Energy Accelerator Research Organization (KEK), Oho, Tsukuba, Ibaraki 305-0801, Japan}
\author{D.V.~Vavilov}\affiliation{Institute for High Energy Physics, Protvino, Moscow Region, 142 280, Russia}
\author{B.~Viren}\affiliation{Brookhaven National Laboratory, Upton, NY 11973, USA}
\author{N.V.~Yershov}\affiliation{Institute for Nuclear Research RAS, 60 October Revolution Pr. 7a, 117312 Moscow, Russia}
\author{Y.~Yoshimura} \affiliation{High Energy Accelerator Research Organization (KEK), Oho, Tsukuba, Ibaraki 305-0801, Japan}
\author{T.~Yoshioka} \altaffiliation{Present address: International Center for Elementary Particle Physics, University of Tokyo, Tokyo 113-0033, Japan.} \affiliation{High Energy Accelerator Research Organization (KEK), Oho, Tsukuba, Ibaraki 305-0801, Japan}
\collaboration{E949 Collaboration}


\newpage

    \begin{abstract}
    A sample of kinematically identified 
    $K^+ \rightarrow \pi^+ \pi^0$ decays obtained with the E949
    detector was used to search for the helicity-suppressed decay
    $\pi^0 \rightarrow \nu \overline{\nu}$ resulting in an upper 
    limit of $2.7\times10^{-7}$ at 90\% confidence level.
    The upper limit is also applicable to $\pi^0$ decays into unknown  
    weakly interacting particles.
    \end{abstract}

\maketitle


We report on a search for the rare decay $\pi^0 \rightarrow \nu \overline{\nu}$
from the E949 experiment~\cite{det.E949,pnnE949,pggE949} 
at Brookhaven National Laboratory (BNL).
The decay is forbidden by angular momentum conservation 
if neutrinos are purely massless left-handed particles.  
A  finite neutrino mass
as evidenced by recent  oscillation measurements  
permits the decay to occur. 
If neutrinos, with mass $m_{\nu}$ less than half of the $\pi^0$ mass, 
couple to the $Z^0$ with standard weak-interaction strength,
the theoretical branching ratio
for the $\pi^0 \rightarrow \nu \overline{\nu}$ decay
is given as
$    Br(\pi^0 \rightarrow \nu \overline{\nu}) = $
        $3\times10^{-8}$ $\left(m_{\nu}/m_{\pi^0}\right)^{2}
        \sqrt{1-4\left(m_{\nu}/m_{\pi^0}\right)^{2}}
$
for a single Dirac-neutrino type~\cite{Robert-equation}.
The experimental upper limit for the tau neutrino mass 
($m_{\nu}$ $<$ 18.2 MeV/$c^2$~\cite{nu-tau-mass})
implies that $Br(\pi^0 \rightarrow \nu \overline{\nu}) < $  $5 \times 10^{-10}$; 
cosmological constraints on the neutrino masses~\cite{WMAPYear1}
imply more stringent limits.  
   The branching ratio for $\pi^0 \rightarrow \nu \nu$ in the 
   case of massive Majorana neutrinos is a factor of two larger~\cite{Marciano-equation}
   than for Dirac neutrinos because the final state particles are identical.
   In addition to $\pi^0 \rightarrow \nu \overline{\nu}$, this search is sensitive to any
   decays of the form $\pi^0 \rightarrow$''nothing''.
   The $\pi^0 \rightarrow$''nothing'' decay can arise from several different
   physics processes beyond the standard model, including
   $\pi^0 \rightarrow \nu \overline{\nu}$ decay
   induced by helicity-flipping (chirality-changing)
   pseudoscalar interactions~\cite{hflip,Prezeau},
   $\pi^0 \rightarrow \nu_{1} \overline{\nu}_{2}$ decay
   where $\nu_{1}$ and $\nu_{2}$ are neutrinos of different lepton flavor,
   and $\pi^0$ decays to other weakly interacting neutral states.
Astrophysical limits on $\pi^0 \rightarrow \nu \overline{\nu}$
have also been adduced from constraints on the cooling of neutron 
stars through the pion-pole mechanism
$\gamma \gamma \rightarrow \pi^0 \rightarrow \nu \overline{\nu}$
\cite{Lam:1991bm}, although nuclear medium effects make this 
model-dependent~\cite{ncool}.

The current upper limit~\cite{pinnE787result} was set by the 
BNL E787 experiment
with $Br(\pi^0 \rightarrow\nu \overline{\nu}) < 8.3\times 10^{-7}$ 
at 90\% confidence level (C.L.)
to all possible $\nu \overline{\nu}$ states.
A flavor specific search for the decay $\pi^0 \rightarrow \nu_{\mu} 
\overline{\nu}_{\mu}$ was performed  by the LSND beam-dump experiment, 
with $Br(\pi^0 \rightarrow \nu_{\mu} \overline{\nu}_{\mu}) < 1.6\times10^{-6}$ 
(90\%C.L.)~\cite{pinnLSND}.

E949 was designed to measure
the rare kaon decay $K^+ \rightarrow \pi^+ \nu \overline{\nu}$~\cite{pnntheory}.
   In that measurement,
   the decay $K^+ \rightarrow \pi^+ \pi^0$ ($K_{\pi2}$)
   is a major potential background 
   and data is analyzed only with $\pi^+$ momenta above \cite{pnnE949,pnnE787} 
   or below \cite{pnn2E787}
   the $K_{\pi2}$ kinematic peak at 205~MeV/$c$.
   In the  $\pi^0 \rightarrow \nu \overline{\nu}$ search, 
   we tag a 205-MeV/$c$ $\pi^0$ in the detector
   by the presence of a $\pi^+$ in the $K_{\pi2}$ kinematic peak. 
   The $\pi^0 \rightarrow \nu \overline{\nu}$ candidates are identified 
   as $K_{\pi2}$ events with no activity other than the $K^+$ and $\pi^+$ 
   in the detector.

An intense beam of 22 GeV/$c$ protons 
from the Alternating Gradient Synchrotron of BNL 
struck a platinum target over a 2.2 s interval 
(spill) every 5.4 s viewed by a beam line~\cite{LESB3} with
two stages of electrostatic mass separation. 
The  typical $K^+$ beam intensity 
(with $K^+$:$\pi^+$ ratio of up to 4:1) at the entrance 
to the E949 detector was $1.3\times 10^7$ per spill with 
momentum 710 MeV/$c$. 
After $K^+$'s were discriminated from $\pi^+$'s by \v{Cerenkov} and energy-loss
counters, they came to rest in a scintillating-fiber target at the rate of 
$3.5 \times 10^{6}$ per spill.  
   The time of the $\pi^+$ that emerged from the target was required to be 
   at least 2~ns later than the time of the incoming $K^+$.
   This ``delayed coincidence'' requirement guaranteed that
   the $\pi^+$ originated from a $K^+$ decay at rest, not from a scattered beam
   particle.
The momenta of the charged decay products were measured 
in a 1 T magnetic field  by a drift 
chamber~\cite{detectorDriftChamber} surrounding the target.
The kinetic energy and range were measured by a cylindrical array of 
plastic scintillators, the range stack (RS), outside of the drift 
chamber. 
The resolutions (rms) of the 
$\pi^+$ momentum ($P_{\pi^+}$), energy ($E_{\pi^+}$) and range ($R_{\pi^+}$)
from $K_{\pi2}$ were 1.1\%, 
2.9\% and 2.9\%, respectively.
Waveform digitizers operating at 500 MHz~\cite{detectorWaveFormDigit} for 
the RS readout recorded the $\pi^+ \rightarrow \mu^+ \rightarrow e^+$ 
decay sequence to distinguish pions from muons.
Photon detectors covered 4$\pi$ sr solid angle to detect any photon or 
extra particle from $K^+$ decay.  A new photon detection device, the 
barrel veto liner (BVL), was introduced just outside the RS and provided 
2.3 radiation lengths to augment the
E787 detector~\cite{detectorE787} configuration; 
with the addition of the BVL, 
a factor of three improvement 
in the $\pi^0$ detection inefficiency 
was expected by Monte Carlo (MC) simulations. 
Additional ancillary photon-detection systems~\cite{Collar} and 
an improved trigger system~\cite{Lev0} were also introduced
into the E949 detector. 
In 2002, the experiment collected 
$N_{K} =$ 
$1.8 \times 10^{12}$ kaons at rest 
in the target in 12 weeks.

The $\pi^0 \rightarrow \nu \overline{\nu}$ search started with the 
identification of $K_{\pi2}$ decays using the $\pi^+$ kinematics 
(``$K_{\pi2}$ tag'') in the events collected by the $K^+ \rightarrow 
\pi^+ \nu \overline{\nu}$ trigger~\cite{Lev0}. Selection criteria 
(cuts) on the $\pi^+$ from the monochromatic two-body decay were set 
at $198 < P_{\pi^+}
< 212$ MeV/$c$, $100 < E_{\pi^+}< 118$ MeV and $28 < R_{\pi^+} < 33$ cm, 
referred to as the ``signal box''.  
Potential non-$K_{\pi2}$ backgrounds include $K^+ \rightarrow \mu^+ 
\nu_{\mu}$ ($K_{\mu2}$) decays and scattered beam pions. These were
suppressed and their contribution to the total background was estimated 
using techniques similar to the $K^+ \rightarrow \pi^+ \nu \overline{\nu}$ 
analysis~\cite{pnnE949}. The $K_{\mu2}$ decays were suppressed with 
measurements of momentum, energy and range as well as with requirements
on the observation of the $\pi^+ \rightarrow \mu^+ \rightarrow e^+$ 
decay sequence. Beam pion background was suppressed by the $K^+/\pi^+$ separation 
in the \v{C}erenkov and energy-loss counters and by requiring the delayed 
coincidence in the target.
Events with two beam particles entering the target, which can defeat the delayed
coincidence requirement, were rejected by looking for activity 
in any of the beam counters at the time of the kaon decay.
The expected numbers of non-$K_{\pi2}$ background events are summarized in 
Table~\ref{table:bifurcation}.  
Ultimately, the search for $\pi^0 \rightarrow \nu \overline{\nu}$ 
was limited by the detection inefficiency 
for the $\pi^0$ decay photons (20--225 MeV) 
from $K_{\pi2}$ decay.

\begin{table}[tb]
    \caption{The number of background and candidate events.} 
    \begin{tabular}{ll}
    \hline \hline
    Total non-$K_{\pi2}$ background            & $3.12^{+1.33}_{-0.99}$   \\
    ~ ~$\bullet$ $K^+ \rightarrow \mu^+ \nu_{\mu}$ (
        $K_{\mu2}$)  & $0.37^{+0.07}_{-0.06}$   \\
    ~ ~$\bullet$ $\pi^+$ beam                  & $0.03^{+0.01}_{-0.01}$   \\
    ~ ~$\bullet$ Two beam particles            & $2.72^{+1.26}_{-0.92}$   \\
    \hline
    Number of $\pi^0 \rightarrow \nu \overline{\nu}$ candidates ($N$)~    & $99$  \\
    \hline \hline
    \end{tabular}
    \label{table:bifurcation}
\end{table}

The single event sensitivity 
for the $\pi^0 \rightarrow \nu \overline{\nu}$ branching ratio $Br$
is given by
\begin{eqnarray}
    SES(\pi^0 \rightarrow \nu \overline{\nu}) 
    &=& \frac{1}{N_{K}Br(K_{\pi2})A_{K_{\pi2}} \cdot C_{dis}C_{acc}} 
    \nonumber \\
    &=& \frac{1}{N_{\pi^0}} \cdot \frac{1}{C_{dis}C_{acc}}
\end{eqnarray}
   where $Br(K_{\pi2})$ is the branching ratio of the $K_{\pi2}$ 
   decay, $A_{K_{\pi2}}$ is the acceptance of the $K_{\pi2}$ tag, and
   $N_{\pi^0}$ is the number of $\pi^0$'s collected by the $K_{\pi2}$ tag.
   A correction factor $C_{dis}$ was introduced 
   to compensate for the loss of $K_{\pi2}$ events from the tagged sample
   due to the misreconstruction of the $\pi^+$ track by overlapping $\gamma$'s and
   $e^{\pm}$'s from the predominant $\pi^0$ decays, which do not occur in the $\pi^0 
   \rightarrow \nu \overline{\nu}$ events.  
   The factor was obtained from two sets of data produced by MC simulations;  
   one was from normal $K_{\pi2}$ 
   decays, and the other was from $K_{\pi2}$ decays where the $\pi^0 \rightarrow
   \nu \overline{\nu}$ decay was forced. The difference in the efficiency of
   the $\pi^+$ reconstruction was used to estimate the correction factor,
   $C_{dis} = 1.14 \pm 0.01$.
   The correction factor $C_{acc}$ takes into account signal losses due to 
   accidental activity in coincidence with the 
   $\pi^0 \rightarrow \nu \overline{\nu}$ decay.
   This factor was obtained from the loss observed in a pure sample of $K_{\mu2}$ decays
   (after all activity of the muons were removed)
   by imposing the cut for hermetic photon detection (HPD).  

\begin{figure}[tb]
    \epsfig{figure=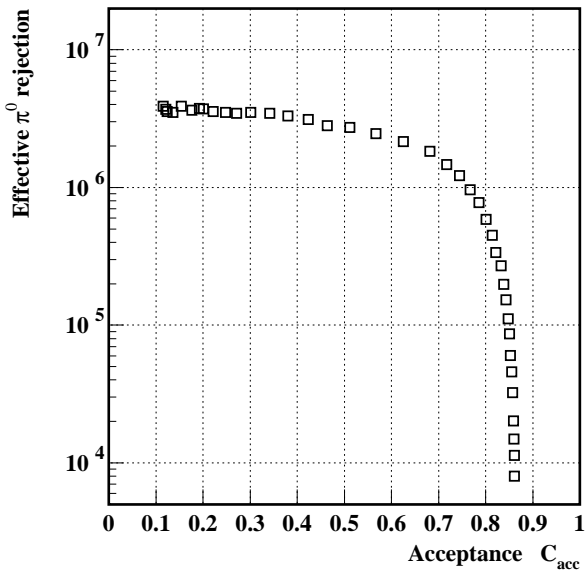, width=8cm}
    \caption{Effective $\pi^0$ rejection (defined as $\pi^0$ rejection 
    $\times$ $C_{acc}$) vs acceptance $C_{acc}$ of the HPD cut
    as measured on the 2/3 sample.
    The saturated curve at $4\times10^6$ indicates the 
    limit of the E949 $\pi^0$ detection efficiency.
    }
    \label{fig:profile-curve}
\end{figure}

   The sensitivity to $\pi^0 \rightarrow \nu \overline{\nu}$ was
   maximized by optimizing the parameters for the HPD cut in order to
   achieve the greatest rejection against the $\pi^0$ decay products 
   ($\gamma\gamma$, $e^+ e^- \gamma$) 
   while minimizing the acceptance loss 
   $(1-C_{acc})$ due to accidentals.
The HPD parameters consisted of timing windows and energy thresholds 
of more than 20 sub-detectors: typically $\pm$10 ns and 1 MeV.
A uniformly sampled 1/3 portion of 
the data (``1/3 sample'') was used as a training sample exclusively
for tuning the parameters. To avoid bias, this sample was not used
for the signal search reported below, nor for the background measurements
shown in Table \ref{table:bifurcation}. After the parameter space
was explored to set the HPD parameters, the cut was imposed on the
remaining 2/3 portion of the data (``2/3 sample'', $N_{\pi^0}=3.02\times10^9$) 
for evaluation. 
The effective rejection (defined as `$\pi^0$ rejection$\times C_{acc}$') 
as a function of $C_{acc}$ for various 
levels
of the cut is shown in
Figure \ref{fig:profile-curve}.
The HPD cut position optimized on the 1/3 sample was set at a value of
$C_{acc}=0.117\pm0.002(stat)\pm0.003(sys)$.
A total of 99 events were observed in the signal box with the final HPD cut. 

\begin{figure}[bth]
    \epsfig{figure=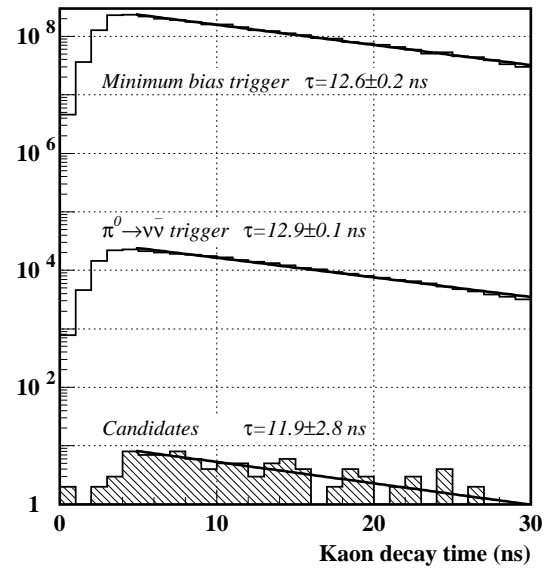, width=8cm}
    \caption{Kaon decay-time distribution with various 
    levels of the HPD cut described in the text. 
    All the other cuts except the offline delayed-coincidence 
    cut were imposed.  The distribution was not distorted by the HPD 
    cut confirming that the sample was dominated by kaon decays. 
    The depletion of events near time zero was due to trigger requirements
    to suppress single beam particle backgrounds.
    Decay-time fits were performed for each plot in a time range of [4ns:30ns]; 
    no evidence of two-beam background was found.
    }
    \label{fig:kaon-life-time}
\end{figure}
\begin{figure}[tbh]
    \epsfig{figure=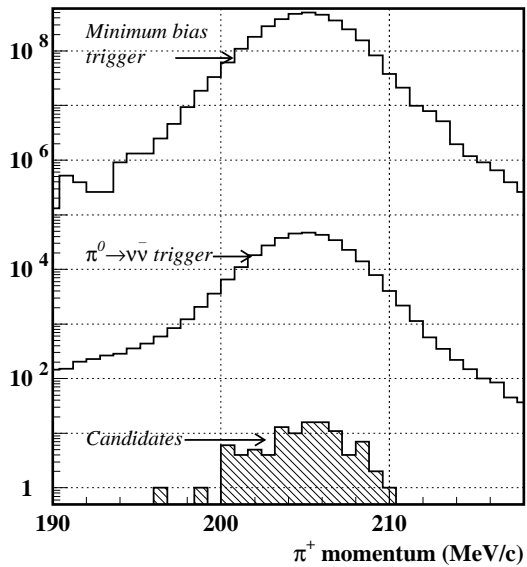, width=8cm}
    \caption{The $K_{\pi2}$ $\pi^+$ momentum distribution
    with various levels of the HPD cut. 
    All the other cuts except for the signal box cuts were imposed.
    }
    \label{fig:kp2-peak-suppress}
\end{figure}

Some properties of the 99 events observed in the signal box are
shown in Figures~\ref{fig:kaon-life-time} and~\ref{fig:kp2-peak-suppress}.
The decay time distribution in Figure \ref{fig:kaon-life-time} is
consistent with the known kaon lifetime and does not show any evidence of
large contamination by two-beam background, which would tend to flatten 
the distribution of the surviving events. 
   The $\pi^+$ momentum distribution in Figure \ref{fig:kp2-peak-suppress} does not show 
   evidence of a significant contamination by beam pions or from $K_{\mu2}$ decays.

The primary reasons for failure to detect photons from $\pi^0$ decay in 
$K_{\pi2}$ events are 
sampling fluctuations in the electromagnetic shower  of low energy photons 
around 20 MeV and 
photonuclear interactions of high energy photons with undetected 
products, such as neutrons. 
While the effects of electromagnetic interactions can be estimated well, 
there are presently very large uncertainties associated with the detailed 
modeling of detection inefficiencies due to  photonuclear processes.  
Therefore, since the overall background 
contribution from $\pi^0 \rightarrow \gamma\gamma$ decays
in which both photons go undetected is difficult to estimate reliably, 
we treated all 99 observed events as $\pi^0 \rightarrow \nu \overline{\nu}$ 
candidates to set an upper limit. Using Poisson statistics, 
the number of signal events was limited to be $<$ 113 at 90\% C.L. when 99 
events were observed.  Subtracting the non-$K_{\pi2}$ background of 
approximately three events, the 90\% C.L. upper limit of the 
$Br(\pi^0 \rightarrow \nu \overline{\nu})$ was obtained as :
\begin{eqnarray}
    Br(\pi^0 \rightarrow \nu \overline{\nu}) &<& 
    \frac{110}{3.02\times10^{9}}\cdot\frac{1}{1.14\times0.117} \\
       && = 2.7 \times10^{-7}
\end{eqnarray}
   The result is three times better than the previous best 
   result~\cite{pinnE787result}. The upper limit obtained above 
   is sensitive to any hypothetical weakly-interacting particles,
   whose masses are less than half of the $\pi^0$ mass;
   other decays of the kind $X^0 \rightarrow$ ``nothing'' 
   (e.g. $\eta$, $K_{L,S}$~\cite{Marciano-equation}, and $B^0$~\cite{BaBarnunu})
   are experimentally more difficult to measure. 

We gratefully acknowledge the dedicated effort of the 
technical staff supporting E949 and of the BNL Collider-Accelerator Department.
We are also grateful to 
R.~Shrock, G.~Pr\'{e}zeau and W.~J.~Marciano for useful discussions 
on the $\pi^0 \rightarrow \nu \overline{\nu}$ decay.
This research was supported in part by the U.S. Department of Energy,
the Ministry of Education, Culture, Sports, Science and Technology of
Japan through the Japan-U.S. Cooperative Research Program in High
Energy Physics and under Grant-in-Aids for Scientific Research, the
Natural Sciences and Engineering Research Council and the National
Research Council of Canada, the Russian Federation State Scientific 
Center Institute for High Energy Physics, and 
the Ministry of Science and Education of the Russian Federation.

\end{document}